\documentclass{iau}

\usepackage{amsmath}
\usepackage{graphicx}
\usepackage{multirow}
\usepackage{natbib}
\begin{document}

\lefttitle{A.I. Ennis}
\righttitle{PNe as tracers of stellar population properties}

\journaltitle{Planetary Nebulae: a Universal Toolbox in the Era of Precision Astrophysics}
\jnlDoiYr{2023}
\doival{10.1017/xxxxx}
\volno{384}

\aopheadtitle{Proceedings IAU Symposium}
\editors{O. De Marco, A. Zijlstra, R. Szczerba, eds.}
 
\title{Planetary nebulae as tracers of stellar population properties:
unlocking their potential with integral-field spectroscopy
}

\author{Ana Inés Ennis$^{1,2}$, Johanna Hartke$^{3,4}$, Magda Arnaboldi$^{5}$, Claudia Pulsoni$^{6}$, Fuyan Bian$^{7}$, Chiara Spiniello$^{8,9}$}
\affiliation{$^{1}$ Waterloo Centre for Astrophysics, University of Waterloo, Canada}
\affiliation{$^{2}$Perimeter Institute, Canada}
\affiliation{$^{3}$Finnish Centre for Astronomy with ESO (FINCA), FI-20014 University of Turku, Finland}
\affiliation{$^{4}$Tuorla Observatory, Department of Physics and Astronomy, FI-20014 University of Turku, Finland}
\affiliation{$^{5}$European Southern Observatory, Garching bei M\"unchen, Germany}
\affiliation{$^{6}$Max-Planck-Institut für Extraterrestrische Physik (MPE), Garching, Germany}
\affiliation{$^{7}$European Southern Observatory, Alonso de Córdova, Vitacura, Santiago, Chile}
\affiliation{$^{8}$INAF – Osservatorio Astronomico di Capodimonte, Salita Moiariello 16, 80131 - Napoli, Italy}
\affiliation{$^{9}$Department of Physics, University of Oxford, Denys Wilkinson Building, Keble Road, Oxford OX1 3RH, UK}

\begin{abstract}
Planetary nebulae (PNe) are essential tracers of the kinematics of the diffuse halo and intracluster light where stellar spectroscopy is unfeasible, due to their strong emission lines. However, that is not all they can reveal about the underlying stellar population. In recent years, it has also been found that PNe in the metal-poor halos of galaxies have different properties (specific frequency, luminosity function), than PNe in the more metal-rich galaxy centers. A more quantitative understanding of the role of age and metallicity in these relations would turn PNe into valuable stellar-population tracers. In order to do that, a full characterization of PNe in regions where the stellar light can also be analysed in detail is necessary. In this work, we make use of integral-field spectroscopic data covering the central regions of galaxies, which allow us to measure both stellar ages and metallicities as well as to detect PNe. This analysis is fundamental to calibrate PNe as stellar population tracers and to push our understanding of galaxy properties at unprecedented galactocentric distances.
\end{abstract}

\begin{keywords}
Catalogs, Surveys, galaxies: elliptical and lenticular, cD, planetary nebulae
\end{keywords}

\maketitle

\section{Introduction}

\begin{figure}
    \centering
    \includegraphics[width=0.6\columnwidth]{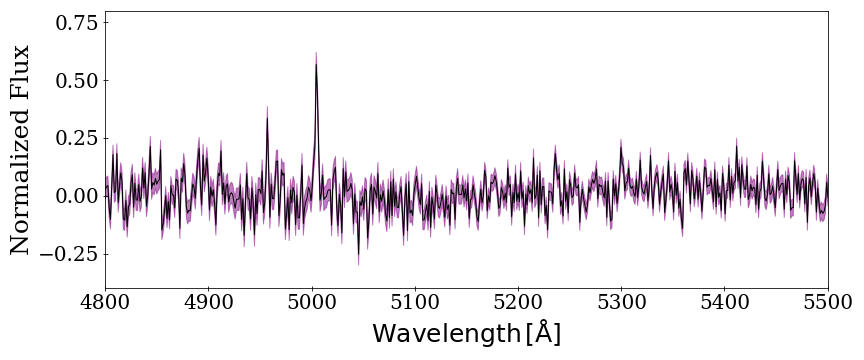}
    \includegraphics[width=0.6\columnwidth]{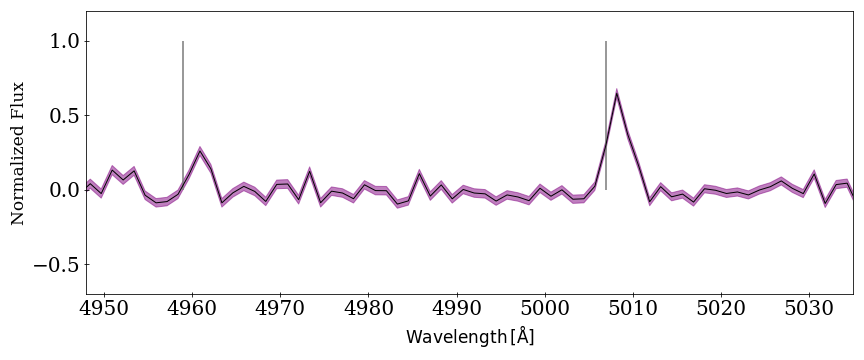}
    \caption{Top: full spectrum for a PNe detected in NGC\,1387. Bottom: zoom-in of the same spectrum in the [OIII] doublet region.}
    \label{fig:spec}
\end{figure}

For almost thirty years planetary nebulae (PNe) have been used as kinematic tracers of the extended haloes of galaxies \citep{Arnaboldi1994,Napolitano2000,Coccato2009,Spiniello2018,Pulsoni2023}. The distinct optical spectral features of PNe, characterized by a flat continuum and prominent emission lines (see Figure\,\ref{fig:spec}), render them detectable at large distances from the centre of galaxies, where the emitted light is too faint for absorption line spectroscopy and stellar population analysis. Furthermore, the luminosity-specific PN number ($\alpha = N_{PN}/L_{gal}$ ratio) has been shown to vary with the age and metallicity of the stellar population of the host galaxy \citep{Buzzoni2006}. A comprehensive understanding of this relationship would unlock the full potential of PNe as tracers of stellar populations, enabling us to characterize age and metallicity in the extended haloes at greater distances than ever before. 

Given that stellar population properties vary throughout the full radial span of a galaxy, it is reasonable to expect these differences may also appear in the PN population if they are connected. Currently, there are very few galaxies for which a joint analysis has been performed of the radial variations of  $\alpha$ and the age and metallicity of its stellar populations. In \cite{Hartke2017}, the $\alpha$-parameter of M49 was shown to vary with colour, presenting lower values in the inner, red, metal-rich halo. Similarly, for M105 in \cite{Hartke2020} two distinct values of $\alpha$ were estimated for different areas of the galaxy, identifying a diffuse outer component with old halo metal-poor stars. Since these studies rely on photometric data, the relation is estimated in terms of colours, which do not translate directly into age and metallicity.

To resolve this degeneracy, it is necessary to analyse PNe in regions of the galaxy where high SNR spectroscopic data of the stellar population can be obtained in order to estimate their age and metallicity independently. A challenge arises, however, as wherever the stellar population is bright enough to allow this, PNe become much harder to detect using traditional methods. The central regions of galaxies, which are perfect candidates for detailed analysis of stellar populations, are usually blind spots for PNe detection techniques.

In this work, we present a way to overcome these limitations by using IFU data from MUSE and the DELF technique (Differential Emission-Line Filter, \citealt{Roth2021}). This enables us to detect PNe while simultaneously analysing the underlying stellar population in the same area. By linking $\alpha$-parameters directly to age and metallicity for a sample of galaxies, we aim to gain a thorough understanding of these relations. This takes us one step closer to PNe becoming reliable tracers of stellar population properties. We aim to perform this analysis on a sample of early-type galaxies which was built using two criteria: that they have MUSE observations publicly available in the ESO Archive, and that they have an available PNe catalogue for their outer halo, for future comparison. We show in these proceedings the results process applied to one of the galaxies in this sample, NGC\,1387.

\section{Detection and classification of sources}

Traditionally, the main detection method for PNe is based on obtaining \textit{on} and \textit{off} images, i.e. a narrowband filter image centered in the [OIII] line, and a broadband filter image which should capture continuum emission \citep{Ford1973,Ford1975,Jacoby1992}. Because PNe have a negligible continuum, they will not show up in the broadband filter, but they will be very bright in the narrow-band one, making them easy to identify among other sources. The main idea behind DELF is that this is easily emulated using IFU data since we can create images based on any filter, and we can enhance our signal by having access to the entire spectrum with high resolution, thus being able to build customised filters in the most convenient way for our purposes.

As a first step, we subtract the continuum emission from the cubes by fitting a polynomial function to each pixel and creating a ''continuum cube", which we then subtract to create an ''emission-line cube" \citep{Spriggs2020}. From the emission-line cube, we extract our \textit{on} images. We create our filter considering that PNe have velocity dispersions of about $\pm 500$\,km/s, so the line can be detected in $5007.8\pm7.5$\AA. Rather than creating a narrowband filter with this width, we split this range into 12 slices with a width of $1.25\AA$, which improves our detection capabilities. We run Source Extractor \citep{Bertin1996} in each of the slices and require sources to be detected in three consecutive slices in order to be considered positive detections. Once we build our catalogue of detected sources, we filter out spurious detections by fitting a Gaussian profile to the flux across all slices. This ensures that we are not detecting extended emission across the entire wavelength range, or potential artifacts in the spectra. Finally, we extract the full spectrum for each source, and perform a visual check on the lines, as seen in Figure\,\ref{fig:spec}.

Since we have access to the full MUSE spectrum for each object, we are able to estimate the flux of multiple emission lines. This makes it possible to construct diagnostic diagrams that separate PNe from other emission-line sources such as supernova remnants and HII regions. We calculate the [OIII] magnitude using the expression from \cite{Jacoby1989}:

\begin{equation}
    m_{\textrm{OIII}}=-2.5\,log_{10}(F_{\textrm{OIII}})-13.74
\end{equation}

\begin{figure}
    \centering
    \includegraphics[width=0.6\columnwidth]{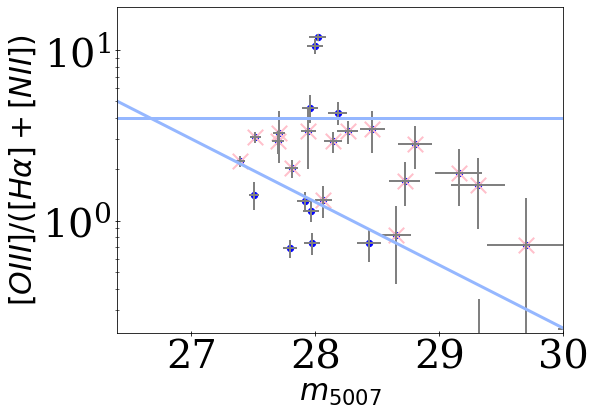}
    \caption{Diagnostic diagram for sources in NGC\,1387. Pink triangles indicate PNe, and blue dots, contaminants. The limits shown in light blue solid lines are defined in \cite{Ciardullo2002}}
    \label{fig:diag}
\end{figure}

In Figure\,\ref{fig:diag} we show the diagnostic diagram for NGC\,1387, where PNe are identified in the region determined by \cite{Ciardullo2002}.

\section{Calculation of the luminosity-specific PNe number}

As a first step, we perform a completeness analysis by injecting artificial stars on the slices using a PSF modeled with a Moffat profile and then repeating the same detection process as before. We compare the fraction of recovered sources to the injected sources in terms of magnitude range and distance to the galactic center.

\begin{figure}
    \centering
    \includegraphics[width=0.6\columnwidth]{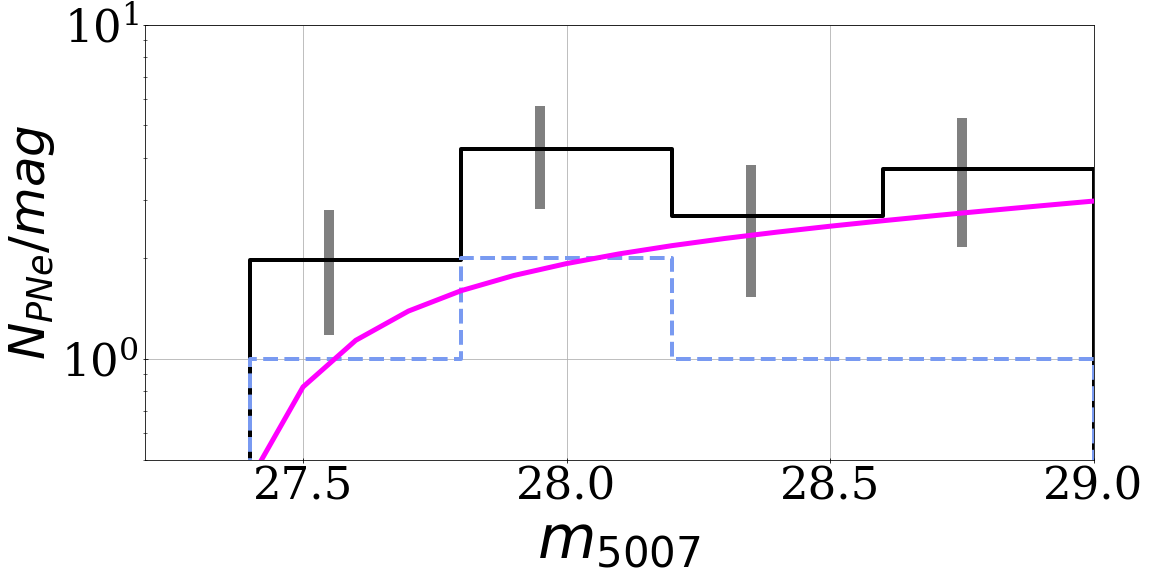}
    \caption{PNLF for NGC\,1387. Dashed blue lines show the original histogram, while grey solid lines show the completeness corrected distribution. The red line shows the function as determined by the distance obtained through SBF method for this galaxy.}
    \label{fig:pnlf}
\end{figure}

We also look into the luminosity function of PNe, which is widely known to be a distance indicator due to the universal absolute magnitude value of its bright cut-off \citep{Ciardullo1989}. Due to the small number of PNe given the limited size of our field of view, we do not aim at measuring distances. However, the PNLF also allows us to estimate what proportion of PNe we are detecting within our completeness limit, and to extrapolate towards fainter magnitudes since its shape has been widely studied. In Figure\,\ref{fig:pnlf} we show the PNLF for NGC\,1387.

\begin{figure}
    \centering
    \includegraphics[width=0.6\columnwidth]{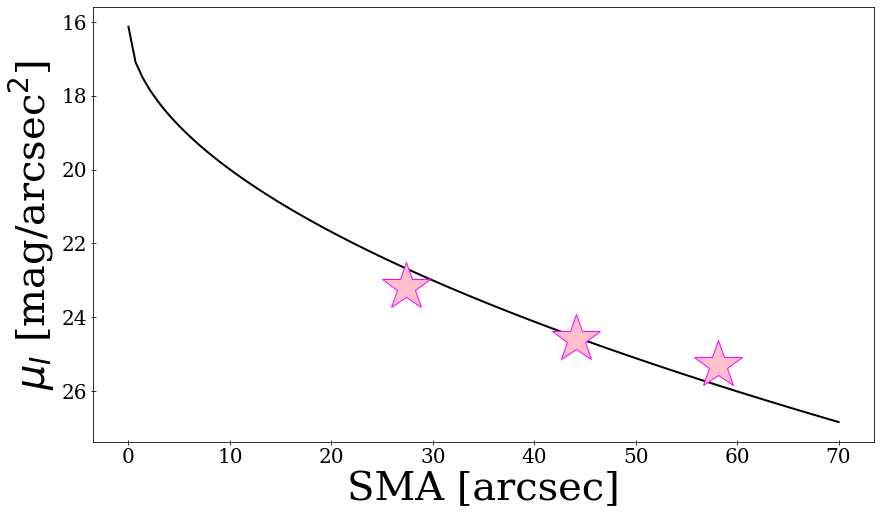}
    \caption{Surface brightness profile of NGC\,1387 in a solid black line, with three pink stars indicating the numeric density of PNe at the corresponding radius, offset by the respective amount.}
    \label{fig:surf}
\end{figure}

The luminosity-specific PN number, $\alpha$, is connected to the total number of PNe, $N_{PN}$, and the total bolometric luminosity $L_{bol}$ through the following expression:

\begin{equation}
    N_{PN} = \alpha L_{bol}
\end{equation}

We bin PNe into elliptical bins, with their geometry defined by the isophotal properties of the host galaxy, and apply the completeness correction. We then measure the PN logarithmic number density profile as 

\begin{equation}
    \mu_{PN}(r)=-2.5 log_{10}\left(\frac{N_{PN,corr}(r)}{A(r)}\right)+\mu_{off}
\end{equation}

where $A(r)$ is the area of the respective ellipse. The offset $\mu_{off}$ is fit to match the surface brightness profile, and we refer to \cite{Hartke2020} for a thorough explanation of how to estimate $\alpha_{2.5}$ from it. For NGC\,1387, we obtain a value of $\alpha_{2.5}=4.39\times10^{-8}\,N_{PNe}\times L_{bol}^{-1}$.

\begin{figure}
    \centering
    \includegraphics[width=0.8\columnwidth]{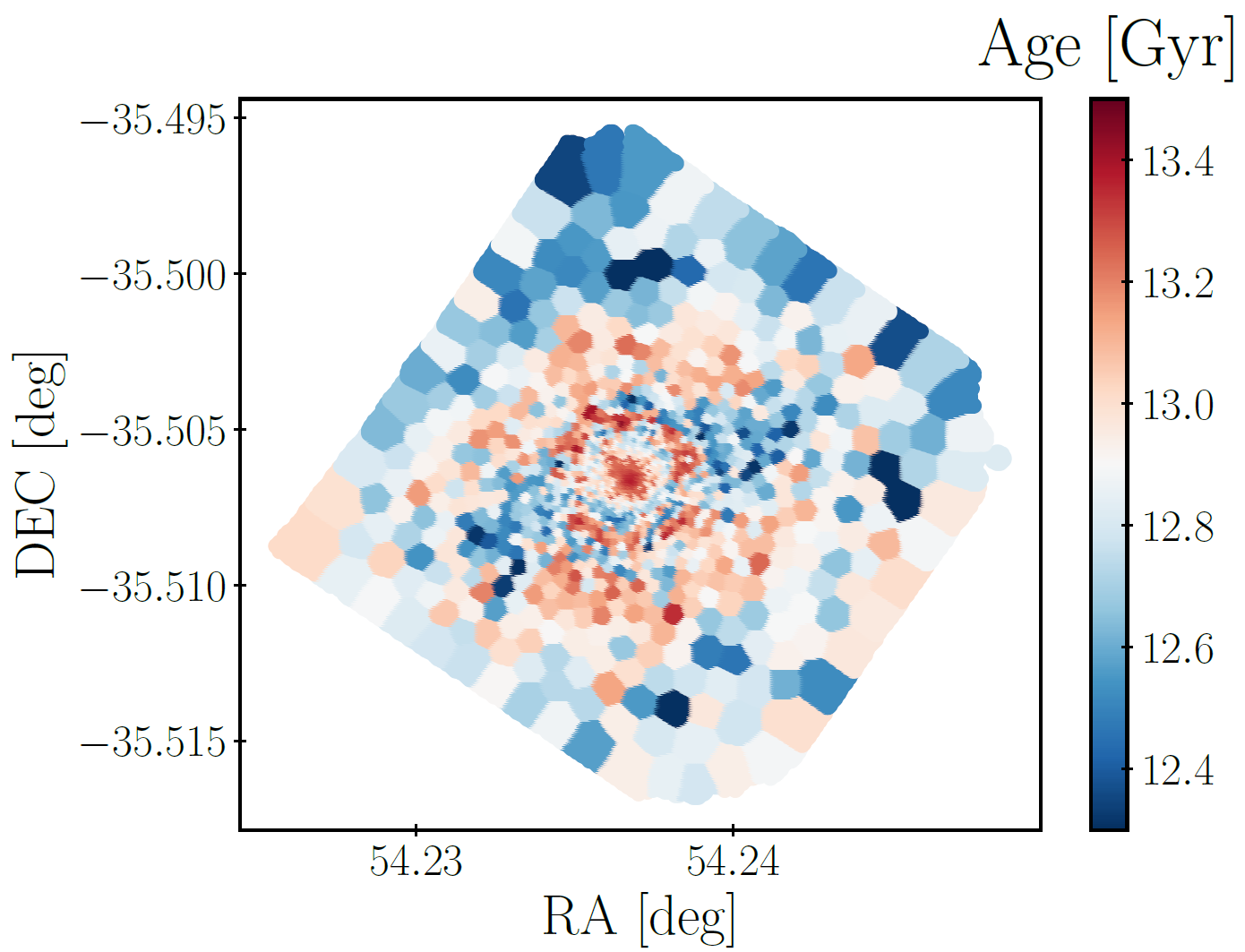}
    \caption{Age estimates for the stellar population of NGC\,1387.}
    \label{fig:kin}
\end{figure}

\begin{figure}
    \centering
    \includegraphics[width=0.8\columnwidth]{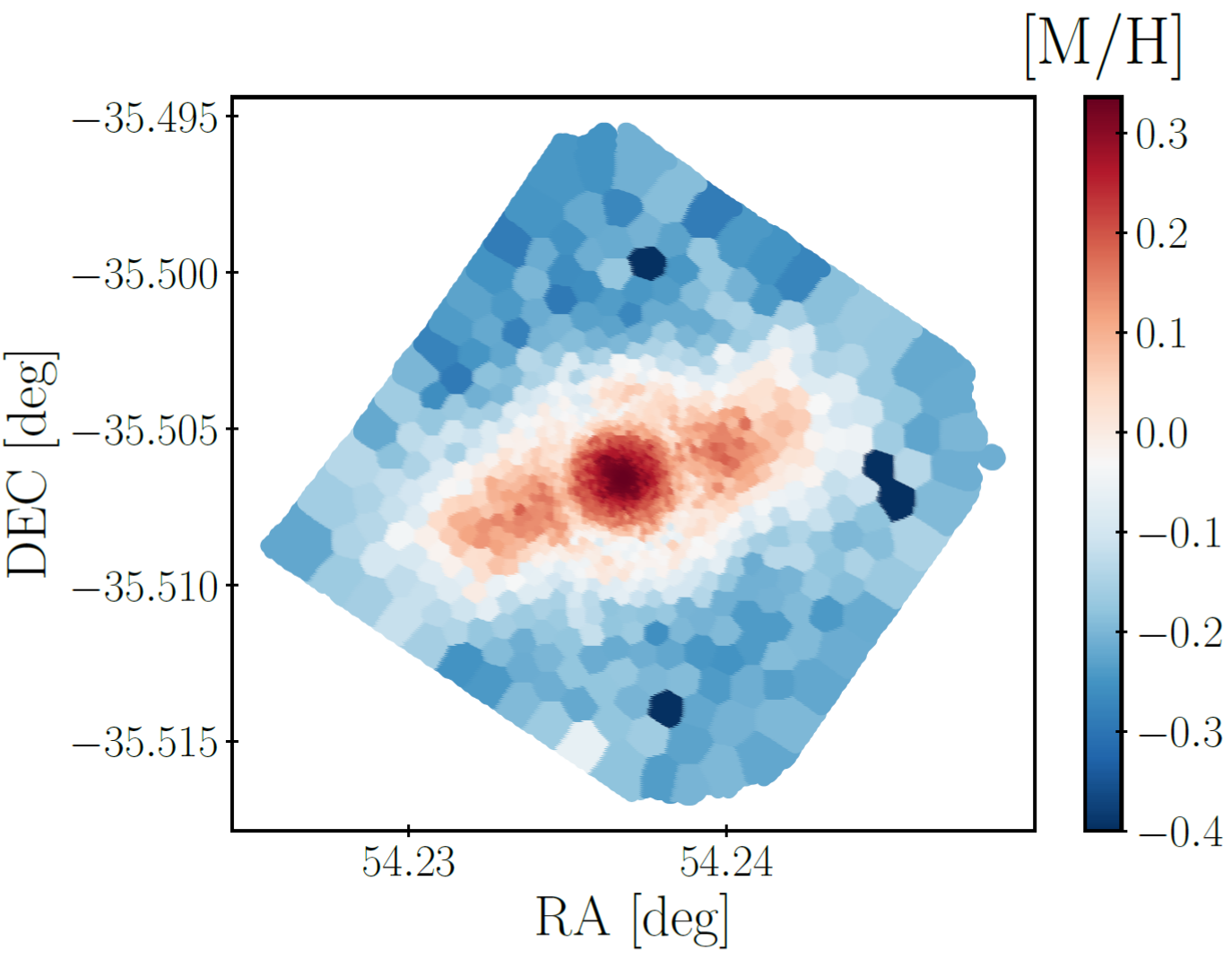}
    \caption{Metallicity estimates for the stellar population of NGC\,1387.}
    \label{fig:sfh}
\end{figure}

We use the GIST pipeline to analyse the stellar population in the cubes. GIST utilizes Penalized PiXel-Fitting  \citep{Cappellari2023} in order to obtain kinematics and stellar population properties (i.e. age and metallicity). In Figs.\,\ref{fig:kin} and \ref{fig:sfh} we show the stellar population properties of NGC\,1387 obtained with this method. 


\begin{figure}
    \centering
	\includegraphics[width=0.8\columnwidth]{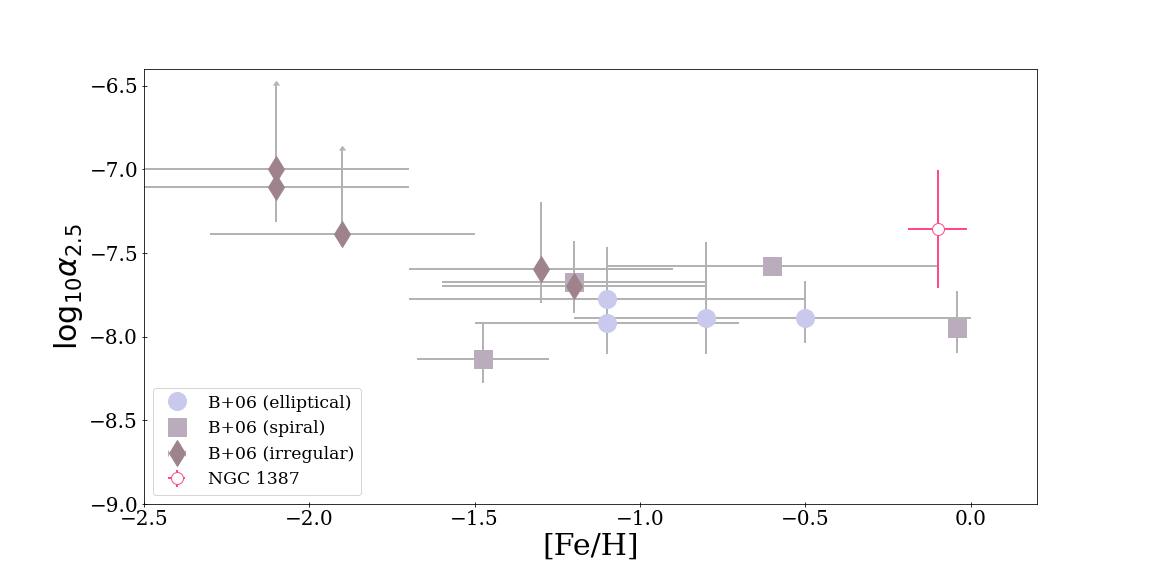}
    \caption{Luminosity-specific PN number of a sample of galaxies from \cite{Buzzoni2006} in relation to their metallicity. The symbols correspond to different morphologies (circles - ellipticals, squares - spirals, diamonds - irregulars). The unfilled circle is our pilot galaxy shown in this work, NGC\,1387.}
    \label{fig:alpha}
\end{figure}

In Figure\,\ref{fig:alpha} we show the relation between $\alpha$ and the mean metallicity of the host galaxy for the sample studied by \cite{Buzzoni2006}, covering a wide range of morphologies, and indicate the position that the inner region of NGC\,1387 has in this relation. Adding the remaining nine galaxies in our sample will allow us to populate the relation, examining how it behaves when the measurements of $\alpha$ and stellar properties are performed in the same regions.

\section{Conclusions}

In this work, we have shown preliminary results for the analysis of PNe in the inner regions of galaxies. We describe the method and show the results for one of the galaxies in our sample. In an upcoming paper, this analysis will be applied to the full sample of ten galaxies, which will enable us to interpret the relation between $\alpha$ and stellar population properties in larger detail.

\section{Acknowledgements}
A.I.E. acknowledges the financial support from the visitor and mobility program of the Finnish Centre for Astronomy with ESO (FINCA), funded by the Academy of Finland grant nr 306531. This research was supported in part by Perimeter Institute for Theoretical Physics.  Research at Perimeter Institute is supported by the Government of Canada through the  Department of Innovation, Science and Economic Development and by the Province of  Ontario through the Ministry of Research and Innovation.

 \bibliography{biblio}

\end{document}